# Ancient bronze disks, decorations and calendars

**Amelia Carolina Sparavigna**
Department of Applied Science and Technology
Politecnico di Torino, Italy

*Recently, it was published that some ancient bronze disks could had been calendars, that is, that their decorations had this function. Here I am discussing an example, the disk of the Trundholm Sun Chariot, proposing a new interpretation of it, giving a calendar of 360 days. Some geometric diagrams concerning the decoration layout are also proposed.*

Some ancient bronze disks, found in burial places in Denmark are covered by amazing decorations. These decorations are composed from several concentric circles and spirals, and bands with zigzag lines. As told in Ref. [1,2], some disks can represent the Sun, which was the supreme power of the Bronze Age cosmology in Denmark. It seems that the religion was based on the daily journey of the Sun and on the progression of the year. It is therefore quite logical to discuss these disks as symbols of time progression and therefore as calendars. This is what is proposed in Ref.2, for some of these items, such as the Trundholn Sun Chariot, a bronze disk and a bronze statue of a horse placed on a device with spoked wheels, and the disk of Egtved [3].

Among the burial objects of the Early Bronze Age, the Trundholm Sun Chariot (Fig.1) is beautiful and amazing for the contrast between the fine decoration of the disk and the stylized shapes of chariot and horse. This artifact is also known as *Solvognen.* The sculpture was discovered in 1902 in a peat bog on the Trundholm moor and is now in the collection of the National Museum of Denmark in Copenhagen. It was cast by the lost wax method [1,4,5]: it means that this technique was known during the Bronze Age. The disk has a diameter of approximately 25 cm. In fact, it consists of two bronze disks, flanged by an outer bronze ring. One of the disks had been gilded on one his side. The models of the disks had been probably decorated with some standard punches, because concentric circles and spirals seem to be identical in the decoration.

The two sides of the disk are considered as representations of the sun, on a chariot pulled by a horse across the heavens from East to West during the day, showing its bright side, the gilded one. During the night, it returns from West to East [1], showing his "dark side" to the Earth. The sculpture is dated to about 1400 BC [1]. However, the same reference is telling, "a model of a horse-drawn vehicle spoked wheels in Northern Europe at such an early time is surprising". They were common in the Late Bronze Age, which ranges from 1100 BC to 550 BC. Ref.1 is suggesting a possible Danubian origin or influence, although the Museum supposes it of Nordic origin.

Let us consider the gilded side of the disk: it has the outer zone, which may represent the solar rays (Fig.2). There is an annulus (the region lying between two large concentric circles) decorated with small multiple concentric circles, linked by a looping band, which creates a "yin and yang" figure (see Fig.3 on the left). The image on the right of the same figure is reproducing the dark side of the sun. In Ref.2 the author is proposing that this side is a calendar. The author, Klaus Randsborg is considering the following calculation. Starting from the centre of the disk, we add the number of spirals in each annulus of the disk, multiplied by the order of the annulus where they are, that is (1×1 + 2×8 + 3×20 + 4×25). This results in a total of 177, a number very close to the number of days in six synodic months. In the Reference 2, the author is also proposing a calendar for the Egtved disk and other objects, supposing that the symbol of "spiral", that is of multiple concentric circles or of a true spiral, is representing the day. The annulus where the symbol is places provides the multiplication factor.

Here I propose another interpretation for the decoration in Fig.3, right panel, that is, of the side corresponding to the night. In the inner part of the disk (see the Figure 4), there are the days of a "week", having therefore 8 days. For the moment, let us not consider the central large circle with many concentric circumferences. It could be a symbol for the cosmos as an ordered and harmonic system, as the cosmos was for the ancient Greeks. In the outer two annuli, there are the weeks of the year, which are 45. Then if we multiply the days in a week by the number of weeks, we obtain 360 days. That is: (8 days) × (45 weeks) = 360 days of the year. As in the ancient Egypt, the year has 360 days: the Egypt divided the year into 12 months of 30 days each, plus five extra days. Let us note that the weeks (see Fig.4) are grouped in two annuli: if we consider the winter solstice as the beginning of the year, the two groups of weeks could have the meaning that during the year there are two seasons, that of a "young sun" followed by the season of a "mature and then old" sun.

Let us note that weeks having eight days existed. The ancient Etruscans developed a week known as the nundinal cycle, around the 8th or 7th century BC. This system passed to Rome, no later than the 6th century BC. It seems that Rome had for a certain period of time a calendar based on two cycles, one having weeks of seven days and the other having eight-day weeks [6]. In any case, using two markers, a marker for the day in the central part and another marker for the week in the outer part, we can use the disk in Fig. 4 as a calendar for a nundinal system. Of course, we need a reference axis, as the black one in the figure. For the five extra days at the end of the year, we can use the circle at the centre of the disk: this is the most important circle, because it is containing both the end and the beginning of the year, able to "adjust" the circle of time, restoring the cosmic order.

For what concerns the other side of the Trundholm disk, the gilded day-side, I can only tell that, if we consider the total number of spirals (52), central included, and assume that each spiral is representing a week having seven days, we can obtain 364. The central "week" is larger because it contains one or two extra days, depending on years. Is it possible that the Trundholm disk is a calendar having two cycles? The answer is beyond the author's knowledges. I consider more reliable the 360 days calendar, as in Fig.4, using the night-side of the disk.

Of course, the decorations in the disk could be simply a beautiful decoration. In any case, if we try to repeat it, we need to arrange in some manner the number of circles/spirals at specific relative distances. The two diagrams of Fig.5 are showing how the artist could have assembled the decoration, subdividing in some angular sectors the disk. It is probable that the artist possessed some specific knowledge of geometric rules. In my opinion, further studies of the decorations of ancient bronze artifacts can be useful to understand the progression of human knowledge of mathematics and geometry.

.

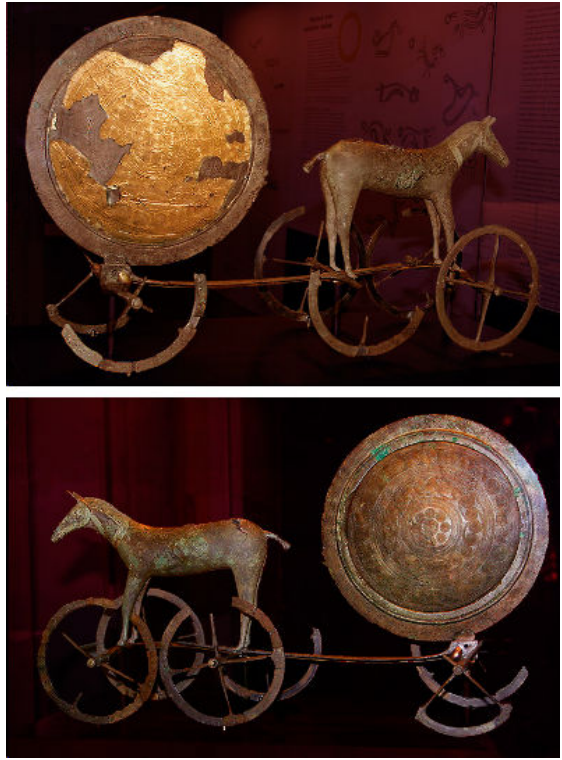

Fig.1 The Trundholm sun chariot (Credit: Malene Thyssen, http://commons.wikimedia.org/wiki/User:Malene)

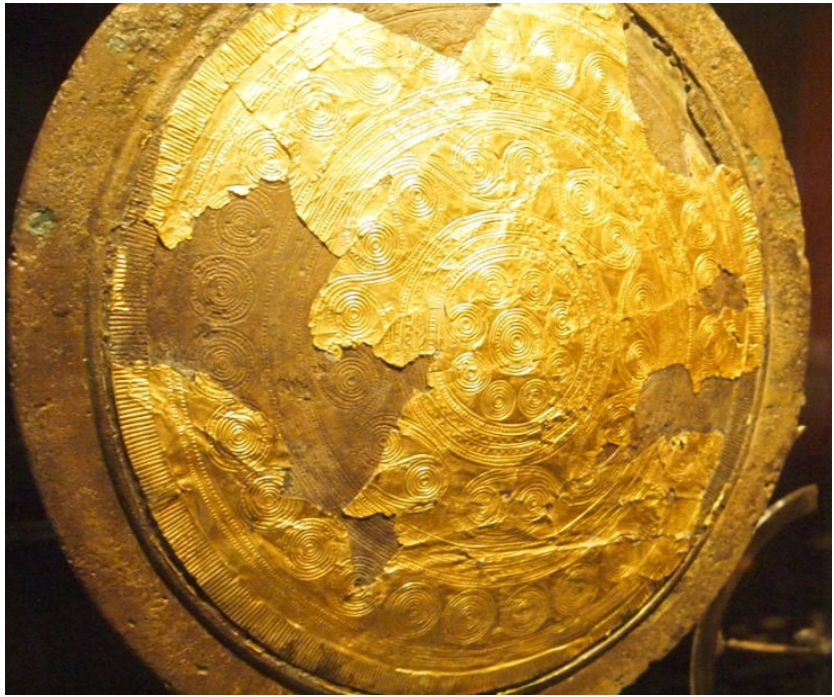

Fig.2 The gilded solar face of the disk. Note the "rays" at the rim of the golden disk (Credit, adapted from a picture taken at the National Museum, Copenhagen Denmark, by Kim Bach).

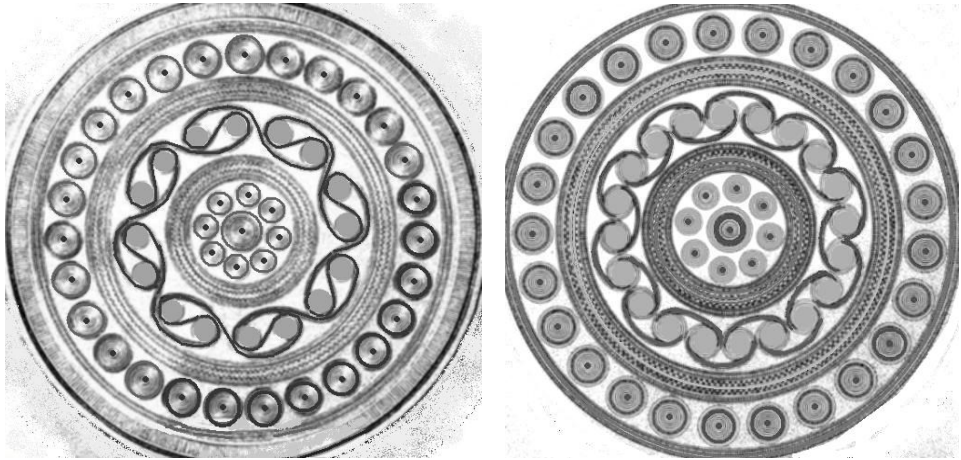

Fig.3 Decorations of the front/day (left) and back/night (right) sides of the disk of the Trundholm sun chariot. Note on the left image, the circles linked by a "yin and yang" pattern. These sketches have been created according to the drawings reported in Reference 2. Please see this reference to see the details, which are amazing

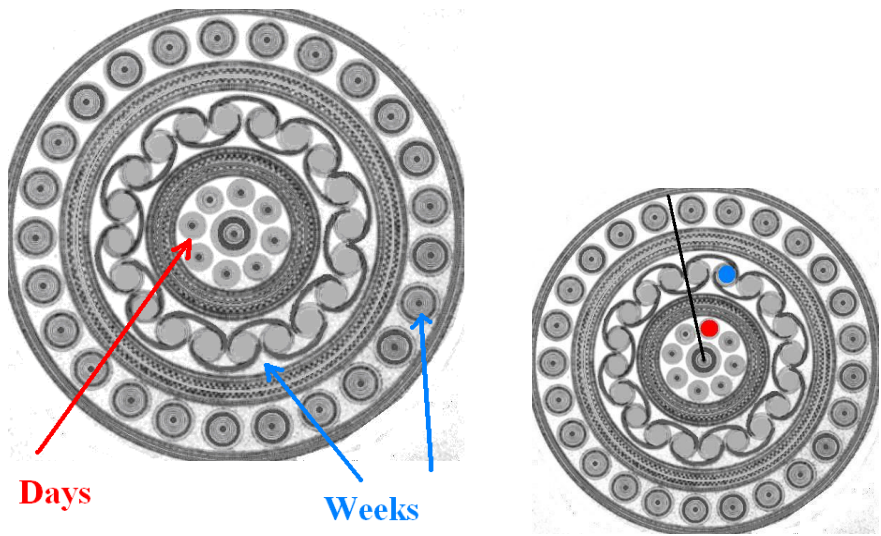

Fig.4. In the inner part of the disk, there are the "days" of a week having 8 days. In the outer two annuli, the weeks of the year, that is 45, subdivided in two "seasons". We have (8 days) × (45 weeks) = 360 days of the year. On the right, it is shown the calendar having the red marker for a day of the week (the first) and a blue marker for a week (the second in the "first season"), according to the reference axis (the black line).

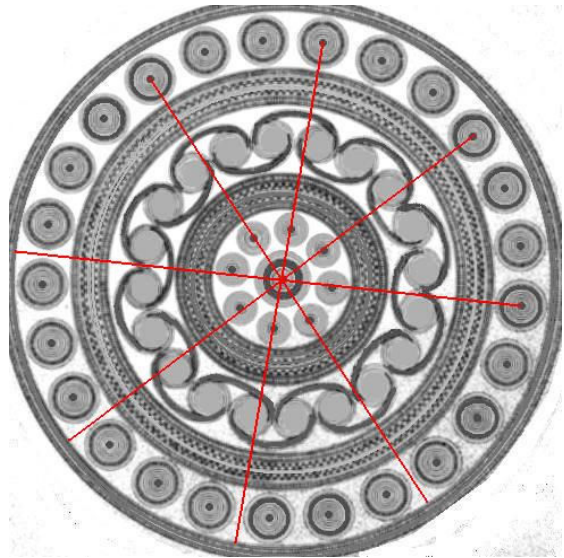

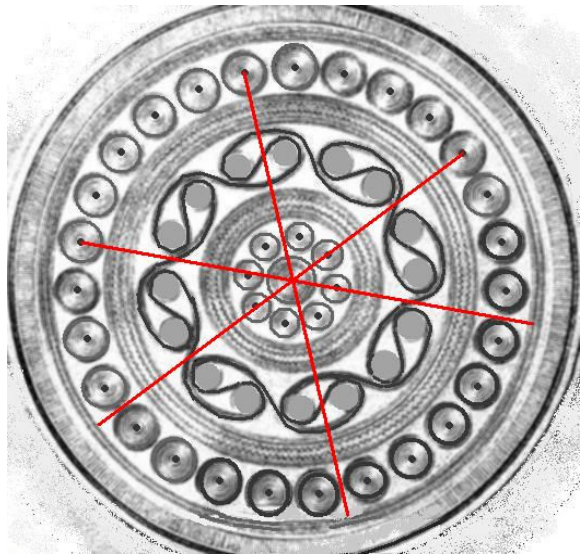

Fig.5. Two diagrams are showing how, probably, the artist had assembled the decoration, subdividing the disk in a few sectors. It seems that the artist knew some geometric rules.